\documentclass[a4paper]{article}

\usepackage{16lomcon}        
\usepackage{cite}             
\usepackage{epsfig}           

\begin{document}


\title{HADRONIC RESONANCE PRODUCTION WITH THE ALICE EXPERIMENT IN pp AND Pb-Pb COLLISIONS AT LHC ENERGIES}

\author{Sergey Kiselev \email{Sergey.Kiselev@cern.ch}
        for the ALICE collaboration
}

\affiliation{Institute for Theoretical and Experimental Physics, 117259
Moscow, Russia}


\date{}
\maketitle


\begin{abstract}
Hadronic resonances $\mathrm{K}^{*}(892)^{0}$, $\phi$(1020) and $\Sigma(1385)^{\pm}$ 
have been measured by the ALICE experiment in pp collisions at $\sqrt{s}=7$~TeV and in
Pb-Pb collisions at $\sqrt{s_{_\mathrm{NN}}}=2.76$~TeV. Transverse momentum spectra, 
particle ratios, nuclear modification factor and comparison with model predictions are discussed. 
In addition, ALICE results are compared with data obtained at RHIC energy.
\end{abstract}

Resonance production plays an important role both in elementary and in heavy-ion collisions.
In pp collisions, it provides a reference for nuclear collisions and also data for tuning 
event generators inspired by Quantum Chromodynamics.
In heavy-ion collisions, the in-medium 
effects related to the highdensity and/or high temperature of the medium can modify the properties 
of short-lived resonances such as their masses, widths, and even their spectral shapes~\cite{Medium}. 
Moreover, due to short life time the regeneration and rescattering effects become important and 
can be used to estimate the timescale between chemical and kinetic freeze-out~\cite{Timescale}.

The resonances have been identified via their main hadronic decay channels: 
$\mathrm{K}^{*}(892)^{0} \rightarrow \pi^{\pm} + \mathrm{K}^{\mp}$, 
$\phi(1020) \rightarrow \mathrm{K}^{+} + \mathrm{K}^{-}$, $\Sigma(1385)^{\pm} \rightarrow \Lambda + \pi^{\pm}$.  
The analysis of the $\mathrm{K}^{*}(892)^{0}$ and $\phi$(1020) mesons in pp collisions at $\sqrt{s}=7$~TeV 
was described in detail in~\cite{ALICEres}.
\begin{figure}[hbtp]
\begin{center}
\includegraphics[scale=0.41]{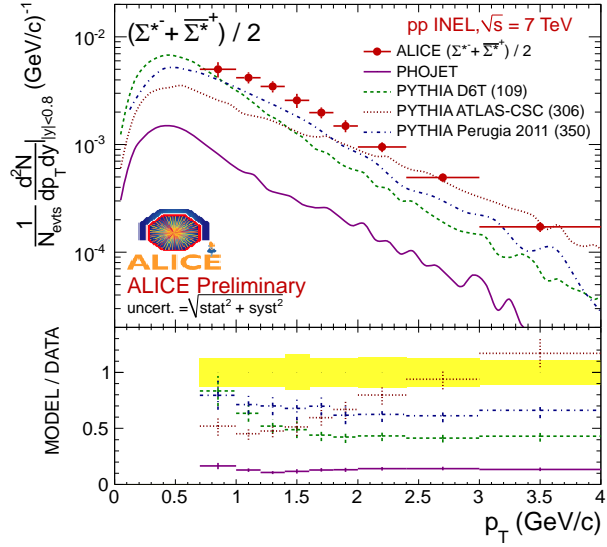}
\end{center}
\caption{Comparison of transverse momentum spectra of  $\Sigma^{*}$ in pp collisions at $\sqrt{s}=7$~TeV 
with PHOJET and PYTHIA tunes D6T, ATLAS-CSC and Perugia 2011.}
 \label{fig:pp-SigmaStar}
\end{figure}
Transverse momentum spectra of the $\mathrm{K}^{*}$ and $\phi$ mesons have been compared 
to a number of PYTHIA~\cite{PYTHIA} tunes and the PHOJET~\cite{PHOJET} event generators. 
None of them is capable of providing a fully satisfactory description of the data.
Figure~\ref{fig:pp-SigmaStar} shows comparison of $\Sigma$(1385) spectra with event generator 
predictions. The models underpredict the data.
The $\mathrm{K}^{*}/\mathrm{K}^{-}$ and $\phi/\mathrm{K}^{-}$ dN/dy ratios 
are found to be independent of energy up to 7~TeV~\cite{ALICEres}. 
Comparing to results from RHIC top energy, $\Sigma^{*}/\pi^{-}$ and $\Sigma^{*}/\mathrm{K}^{-}$ ratios are also independent of energy within
errors and agree with the thermal model~\cite{Becattini} predictions at $\sqrt{s}=7$~TeV. 
For the $\Sigma^{*}/\Xi^{-}$ ratio there is a hint of decrease with energy and its value at $\sqrt{s}=7$~TeV
is overpredicted by the thermal model~\cite{BaldinXXI}.

\begin{figure}[hbtp]
\begin{center}
\includegraphics[scale=0.41]{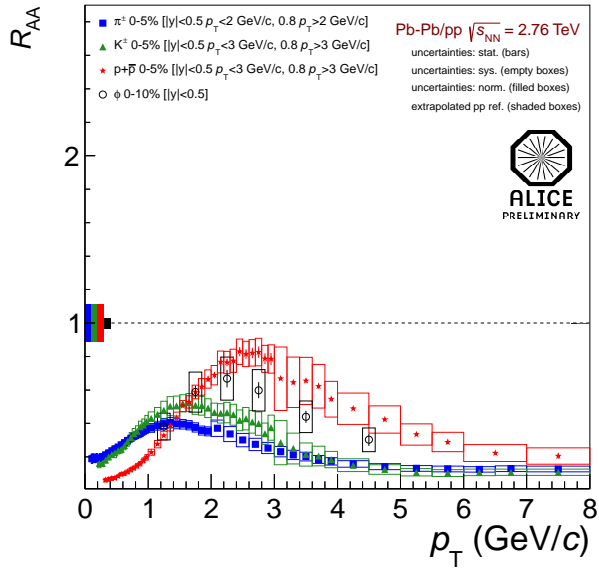}
\end{center}
\caption{\ensuremath{R_{\mathrm{AA}}} of various particles.
}
  \label{fig:PbPb-ratios-RAA}
\end{figure}

\begin{figure}[hbtp]
\begin{center}
\includegraphics[scale=0.50]{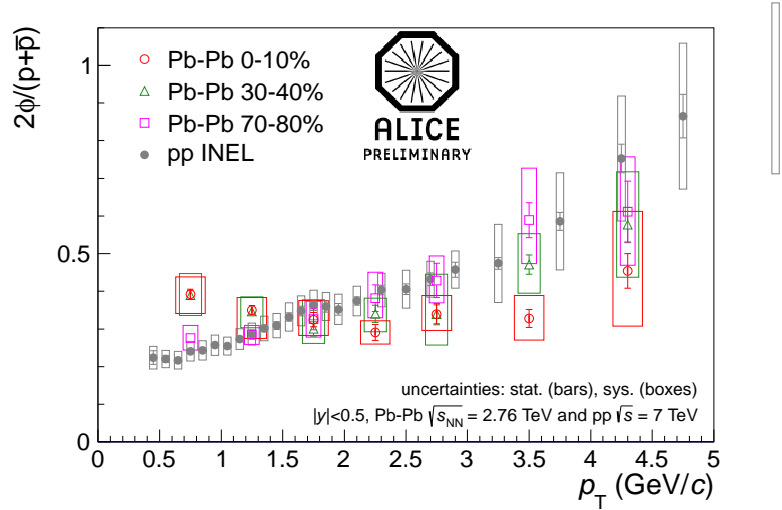}
\end{center}
\caption{$\phi/p$ ratio.}
  \label{fig:PbPb-ratio2pVsPT}
\end{figure}

The mesonic resonances $\mathrm{K}^{*}$ and $\phi$ have been also measured
in Pb-Pb collisions at $\sqrt{s_\mathrm{NN}}$ = 2.76 TeV.  
For $\mathrm{K}^{*}$($\phi$) the $\langle p_\mathrm{T}\rangle $ at LHC energy is $\sim20\%$($\sim30\%$)
higher than that observed at RHIC energy~\cite{NN2012}.
A weak centrality dependence was observed in the $\mathrm{K}^{*}/\mathrm{K}^{-}$ ratio, while the $\phi/\mathrm{K}^{-}$ 
ratio is independent of the collision centrality~\cite{CPOD13}. The decreasing trend in the
$\mathrm{K}^{*}/\mathrm{K}^{-}$ ratio suggests 
a possible increase of hadronic rescattering in the most central collisions. 
The nuclear modification factor \ensuremath{R_{\mathrm{AA}}} for four different particle species is shown in 
Figure~\ref{fig:PbPb-ratios-RAA}.
For $p_\mathrm{T}<2.5$ GeV/$c$, the $\phi$ \ensuremath{R_{\mathrm{AA}}} appears to follow \ensuremath{R_{\mathrm{AA}}} of p, while for mid-to-high $p_\mathrm{T}$, the $\phi$ \ensuremath{R_{\mathrm{AA}}} tends to be between \ensuremath{R_{\mathrm{AA}}} of the mesons ($\pi$ and K) and p. 
The $\phi/p$ ratio as a function of $p_\mathrm{T}$
demonstrate a flat shape for central collisions, Figure~\ref{fig:PbPb-ratio2pVsPT}.
That indicates that the differences in shape between the $\phi$ and p \ensuremath{R_{\mathrm{AA}}} 
are most likely due to the pp reference.  
The ratio for peripheral collisions is similar to pp at $\sqrt{s}=7$~TeV.

\end{document}